\documentclass[10pt]{article}
\usepackage{latexsym}
\usepackage{amsfonts}
\usepackage{graphicx}
\usepackage{amssymb}
\usepackage[numbers]{natbib}
\usepackage{subfig}
\usepackage{tikz}
\usepackage{multicol}
\usetikzlibrary{shapes.symbols,shapes.geometric,shadows,arrows.meta}
\tikzset{>={Latex[width=1.5mm,length=2mm]}}
\usepackage{flowchart}\usepackage[paperheight=11.69in,paperwidth=8.26in,left=0.61in,right=0.61in,top=0.61in,bottom=0.61in,headheight=1in]{geometry}
\usepackage[utf8]{inputenc}
\usepackage[T1]{fontenc}
\usepackage{helvet}

\usepackage[font=small,labelfont=bf]{caption}

\newenvironment{Figure}
  {\par\medskip\noindent\minipage{\linewidth}}
  {\endminipage\par\medskip}

\usepackage{tikz}
    \usetikzlibrary{shadows}
\usepackage{tcolorbox}
    \tcbuselibrary{skins}

\definecolor{esoblue}{HTML}{0070C0}
\newcommand{\mytitle}[1]{
    \node[fill=white,
        draw=esoblue,
        line width=0.5pt,
        text width=1.75cm,
        inner sep=8pt,
        xshift=-3.7cm]
    at (frame.north){\bfseries\textcolor{black}{#1}};
}

\newtcolorbox{mybox}[2][]{
    enhanced,
    overlay={\mytitle{#2}},
    borderline={.7pt}{0mm}{esoblue},
    frame hidden,
    arc=0mm,
    sidebyside,
    lefthand width=2.5cm,
    segmentation hidden,
    top=15pt,
    #1
}

\newcommand{\arttitle}[1]{\fontsize{24pt}{32pt}\selectfont \textcolor[HTML]{0070C0}{#1}\par}
\newcommand{\artauth}[1]{\fontsize{12pt}{18pt}\selectfont \textbf{#1}\par}
\newcommand{\artaff}[1]{\fontsize{11pt}{16pt}\selectfont #1 \par}

\newcommand{\TheTitle}{The Impact of the CMB on high-$z$ AGNs jets}
\newcommand{\MyName}{Luca Ighina}
\newcommand{\MyInst}{Insubria University (Como, Italy) + INAF--Brera (Milano, Italy)

}

\begin{document}

\begin{center}
\arttitle{\TheTitle}\par
\artauth{\MyName}\par
\artaff{\MyInst}
\end{center}\par

\addcontentsline{toc}{section}{{\MyName} - {\it \TheTitle}}
\vspace{0.5cm}
\begin{multicols}{2}


\noindent 
Radio-Loud (RL) Active Galactic Nuclei (AGNs) are among the brightest astrophysical sources at all wavelengths. Their relativistic jets can affect both their Supermassive Black Holes (SMBHs) growth and the surrounding intergalactic medium. While in the radio band these jets can be observed at all scales (from pc to Mpc scales), their X-ray and $\gamma$-ray emission is expected to be concentrated on very small scales (\textless10~pc). However, after the launch of the Chandra X-ray telescope, several kpc-scale jets were detected \citep{schwartz99} and the mechanism responsible for their high-energy radiation at these scales is still under debate. Understanding its origin is crucial also to derive the physical properties of these jets (e.g. the power) at large scales and, as a consequence, their impact on the  environment.
In the following, we explore the Inverse Compton interaction of the relativistic electrons within relativistic jets with the Cosmic Microwave background photons (IC/CMB) as possible interpretation. Moreover, we also estimate how this interpretation could also affect the observed evolution across cosmic times of the SMBHs hosted in jetted systems.

\vspace{0.25cm}
{\fontsize{10pt}{10.8pt}\selectfont \textcolor[HTML]{0070C0}{The most distant large-scale relativistic jet, at $z=6.1$}\par}
\noindent After the detection of several large-scale jets with the Chandra telescope, one of the most popular interpretations proposed for their X-ray emission was that it is produced by the IC/CMB interaction \citep{tavecchio00}. Despite requiring quite extreme assumptions on the physical parameters (e.g. bulk Lorentz factor of the kpc-scale jets $\Gamma\sim$10-20), this model was successful in reproducing the high X-ray fluxes observed in the majority of these large-scale jets. However, after the launch of the Fermi $\gamma$-ray telescope, this model was challenged by the non-detection of the expected strong $\gamma$-ray emission in a few objects in the low-$z$ Universe ($z<1$). Therefore, another interpretation was proposed, with the X-ray radiation produced by a second population of highly-energetic electrons (with Lorentz factors $\gamma\sim10^{8-9}$) through synchrotron (see \citep{georg16}). Nevertheless, we still expect the IC/CMB process to take place, especially at high redshift. Indeed, given the strong redshift dependence of the CMB energy density, $U_{\rm CMB}\propto(1+z)^4$, and that the IC/CMB is significantly boosted for jets oriented close to our line of sight (i.e. for sources classified as blazars), the best way to constrain it is to focus on high-$z$ blazars. To this end we performed a Chandra X-ray observation of the most distant blazar currently known, PSO~J0309+27 at $z=6.1$ \citep{belladitta20}. This observation revealed the presence of a $\sim$20~kpc jet extending in the North-East direction (see Figure \ref{IghinaFig1}), making it the highest-redshift large-scale jet currently known. From the detailed study of its multiwavelength emission (see \citep{ighina22}), we found that its X-ray flux is fully consistent with the IC/CMB process with $\Gamma \sim2$, therefore without requiring the extreme bulk velocities velocities needed at low redshift. This is due to a double effect of the CMB energy density increase at high redshift: (a) it boosts the IC/CMB process by a factor $(1+z)^4$; (b) it decreases the cooling timescale of the electrons, which, in turn, cannot be accelerated up to the very high energies needed to emit in the X-rays through synchrotron. Therefore, from our work we concluded that, while in the local Universe the synchrotron process likely dominates the X-ray emission observed in large-scale relativistic jets, at high redshift ($z\gtrsim2$) the IC/CMB process becomes the most important mechanism.
\begin{Figure}
\centering
\includegraphics[width=0.87\linewidth]{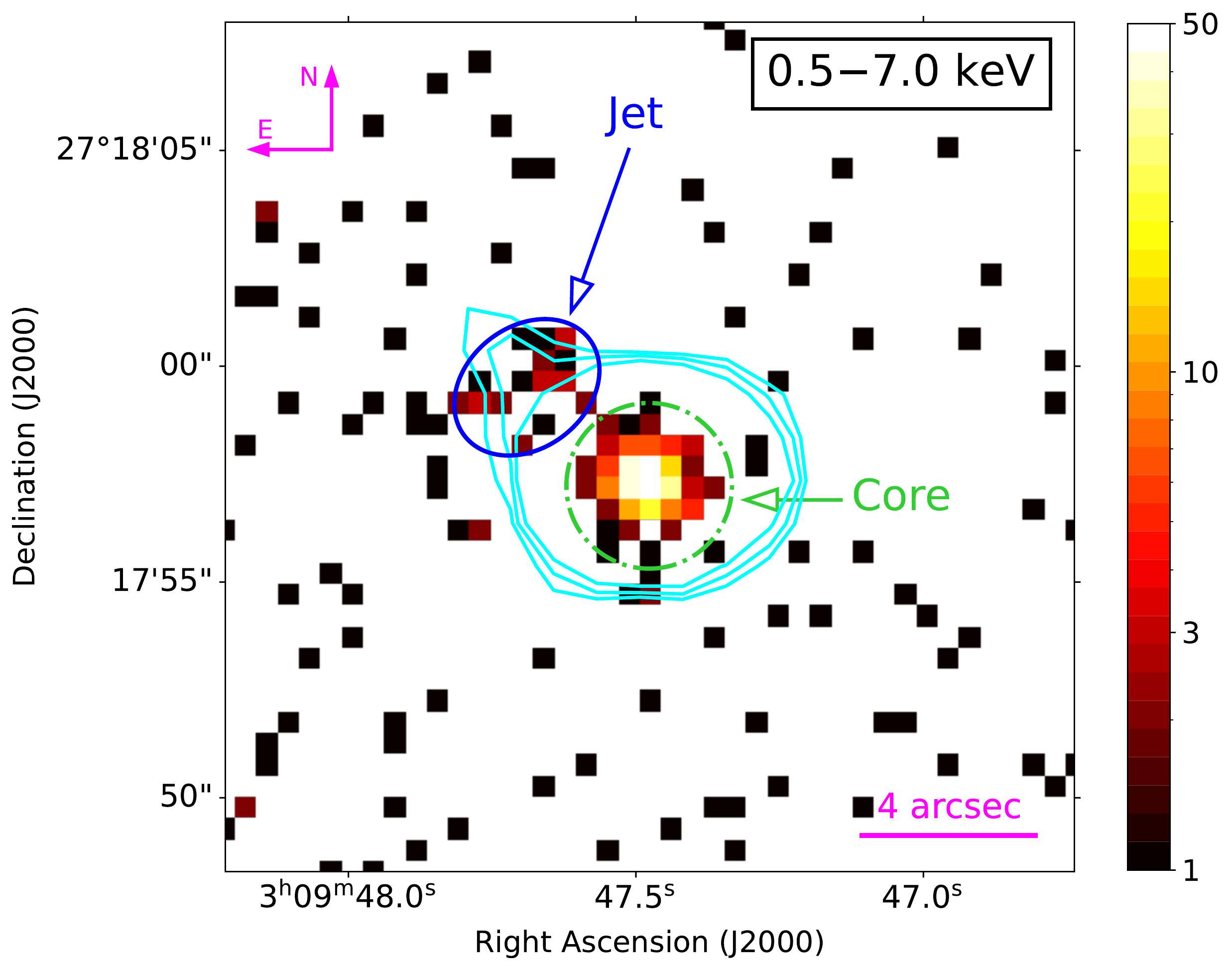}  
\captionof{figure}{\small Chandra X-ray image of PSO~J0309+27 with the (3, 3$\sqrt{2}$, 6)$\times$RMS radio contours at 3~GHz from the VLASS survey overlaid. The kpc-scale jet is highlighted by the blue ellipse. Adapted from \citep{ighina22}.} 
\label{IghinaFig1}
\end{Figure}
\vspace{0.5cm}

{\fontsize{10pt}{10.8pt}\selectfont \textcolor[HTML]{0070C0}{Space density evolution of blazars}\par}
\noindent 
If, as explained above, the IC/CMB process was important and significant at high redshift, we would expect the overall X-ray emission (core plus extended jet emission) of blazars to increase as a function of redshift. Interestingly, such a trend was found from the comparison of statistically complete samples of blazars at high and low redshift (e.g. \citep{ighina19}). In particular, these studies found that $z\sim4.5$ blazars have, on average, X-ray-to-radio luminosity (X/R) ratios about two times larger than their low-$z$ counterparts. This difference could also have important consequences on our understanding of the evolution of blazars, and therefore of the jetted SMBHs, through cosmic times. Indeed, recent studies have found that the cosmological evolution of the blazar population significantly differs when observed in the X-ray or the radio band. In particular, X-ray studies (e.g. \citep{ajello09}) find that the number of blazars at high redshift is significantly larger than the one expected from the radio band (e.g. \citep{mao17}). As shown also in Figure \ref{IghinaFig2}, the X-ray space density evolution of blazars, computed starting from the radio Luminosity Function (LF) derived by \citep{mao17} and assuming a constant X/R conversion factor, under-predicts the number of blazars compared to the observed X-ray data (green points). However, if we now consider that a fraction of the overall X-ray emission is due to the IC/CMB process, we expect the typical X/R ratios of blazars to evolve with redshift as follows:\\

\noindent 
L$_{X}$/L$_{r}$~($z$)~=~L$_{X}$/L$_{r}$ ($z=0$)~$\times$~($1-A_0$)~+~$A_0~\times~(1+z)^4$,\\ 

\noindent 
where $A_0$ is the relative contribution of the IC/CMB emission at $z=0$. Based on the study of the largest complete samples of blazars selected in the radio band, we found $A_0$ to be of the order of $\sim 10^{-3}$, that is, the IC/CMB contribution at low redshift is negligible, as expected. By including this type of X/R evolution in the computation of the expected X-ray space density starting from the radio LF, we found that the increase of the X-ray emission related to the IC/CMB process can reconcile the evolutions observed in the radio and X-ray band, see the solid red curve in Figure \ref{IghinaFig2}.\\
In order to improve the current statistics at high redshift and therefore to further test the IC/CMB mechanism, the e-ROSITA X-ray mission will be crucial, thanks to its relatively deep sensitivity over the entire sky. Indeed, the actual number and redshift distribution of the blazars detected by this mission will depend on the importance of the IC/CMB process to the observed X-ray emission. At the same time, detailed studies on these X-ray-selected sources with the Chandra telescope, like the one presented above, will further constrain the IC/CMB model.

\begin{Figure}
\centering
\includegraphics[width=0.87\linewidth]{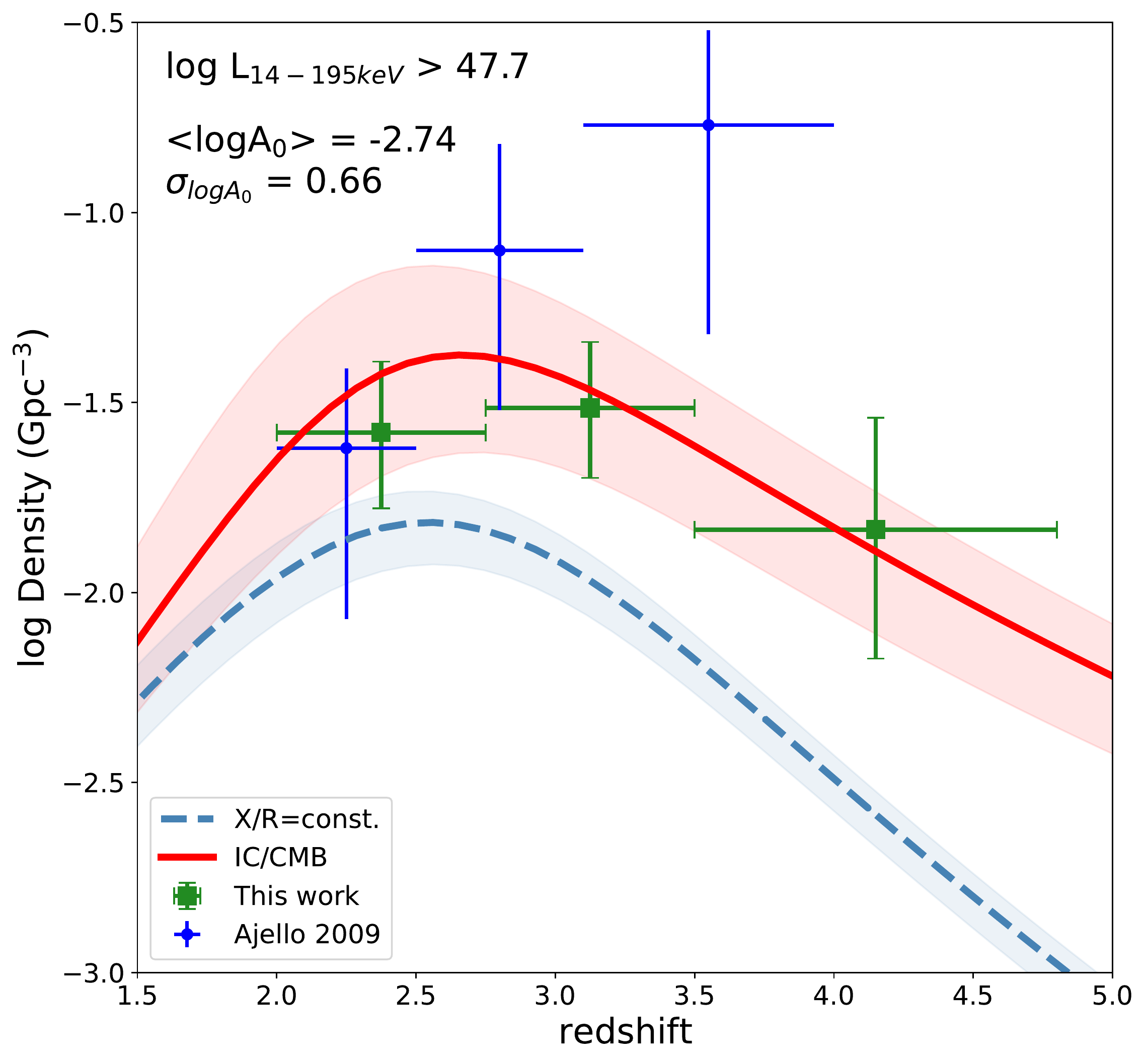}  
\captionof{figure}{\small Observed X-ray space density of blazars: green squares from \citep{ighina21b}, blue points from \citep{ajello09} as reference. The two lines are the expected space density evolution of blazars with (solid red) and without (dashed blue) the redshift evolution given by the the IC/CMB process. Adapted from \citep{ighina21b}.}  
\label{IghinaFig2}
\end{Figure}
\vspace{0.5cm}

{\fontsize{10pt}{10.8pt}\selectfont \textcolor[HTML]{0070C0}{Hunting down RL AGNs at high redshift}\par}
\noindent  If, on the one side, the e-ROSITA mission will uncover the X-ray-brightest end of the blazar population, on the other side, on-going and future radio surveys, like the ones performed by the Square Kilometre Array Observatory and its precursors, will uncover the much larger RL population of AGNs at high redshift. Indeed, soon after the first data release of the Rapid ASKAP Continuum Survey (RACS) we were able to identify one of the most distant RL AGNs currently known: VIK~J2318$-$3113, at $z=6.44$ \citep{ighina21a}. Given this first successful identification, we are now using the RACS survey as starting point for the discovery of new RL AGNs at high redshift. In particular, from its combination with optical/NIR wide-area surveys currently available (namely Pan-STARRS and DES) and the use of the \textit{dropout} technique we are building the first statistically complete sample of RL AGNs at $5<5<6.5$. In Figure \ref{IghinaFig3} left panel we report the sky coverage of the RACS and the optical/NIR surveys used together with the sky distribution of the currently known $z>5$ RL AGNs and our newly selected candidates. While the spectroscopic confirmation of the majority of the candidates is still ongoing, we were already able to identify two new $z>6$ RL AGNs through dedicated Gemini--South/GMOS observations (GS-2021-DD-112, P.I. Ighina; see Figure \ref{IghinaFig3} right panel). Chandra observations of these sources will potentially allow us to detect and study more kpc-scale jets at $z>6$ to further test the IC/CMB model.

\begin{Figure}
\centering
\includegraphics[width=0.95\linewidth]{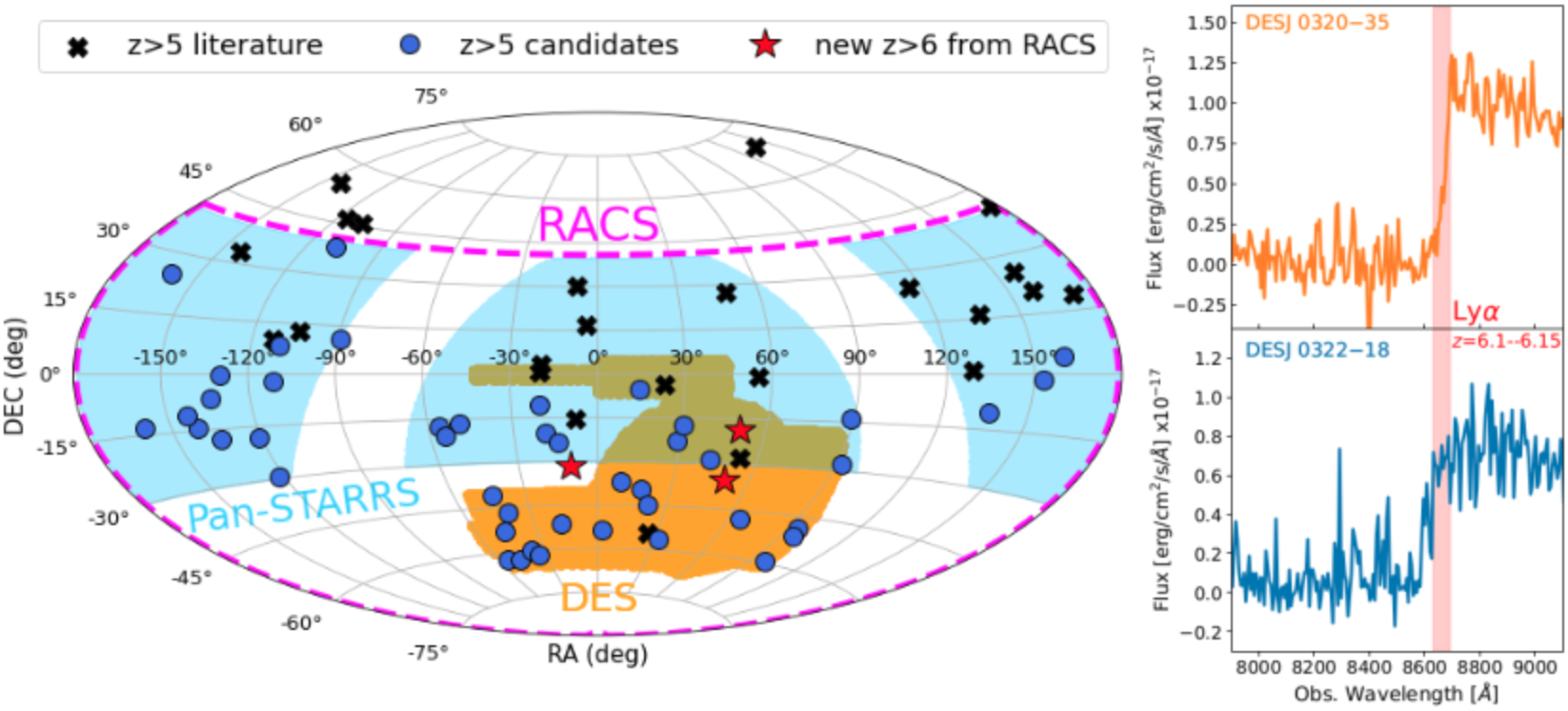}  
\captionof{figure}{\small Left panel: Sky coverage of the radio and optical/NIR surveys used for our selection together with the currently known, the new selected candidates and the newly identified RL AGNs at $z>5$. Right panel: zoom on the optical/NIR dropout of the two newly confirmed RL AGNs at $z>6$.}  
\label{IghinaFig3}
\end{Figure}
\vspace{0.5cm}

{\fontsize{9pt}{9.8pt}
\selectfont \textcolor[HTML]{0070C0}{References}\par 

\begingroup
\renewcommand{\section}[2]{}%

\endgroup
}
\end{multicols}

\vspace*{\fill}
\begin{mybox}{Short CV}
    \includegraphics[scale=.095]{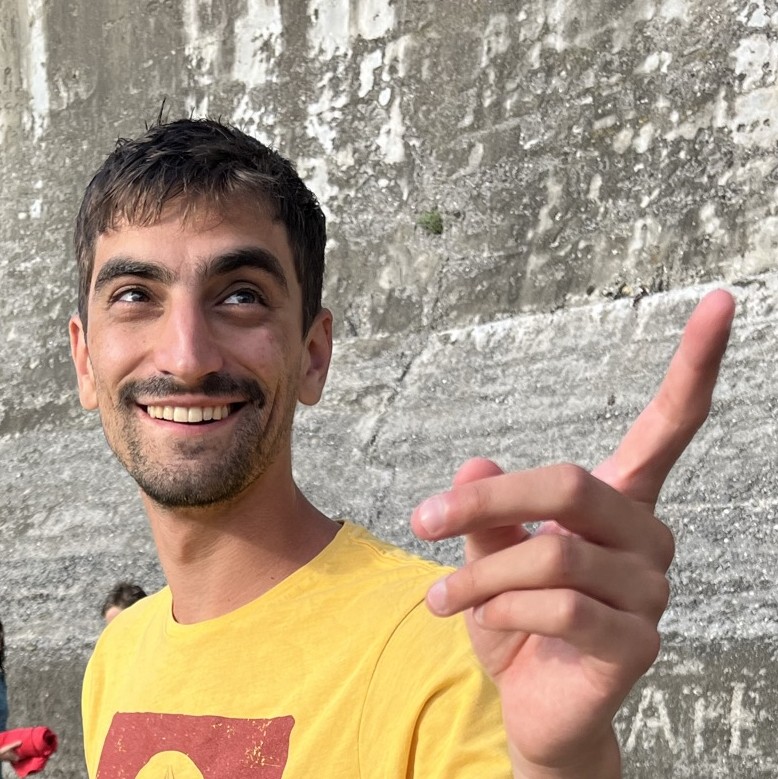}   
    \tcblower
    \begin{tabular}{l@{\hspace{1\tabcolsep}}l}
    2018--2020:& MSc in Astrophysics, Milano--Bicocca University, Milano, Italy\\
    2020--present:& PhD in Physics and Astrophysics, Insubria University, Como, Italy\\
    2024--on: & Relaxing on a beach somewhere in the Caribbeans 
    \end{tabular}
\end{mybox}

\end{document}